\documentclass[a4paper,12pt]{article}
\usepackage{epsfig}
\newcommand{\vp}{\vec{p}}
\newcommand{\vlamc}{\vec{\Lambda}_c^+}
\newcommand{\vg}{\vec{\rm g}}
\newcommand{\vc}{\vec{c}}
\newcommand{\vzero}{\vec{0}}
\newcommand{\svp}{\vp}
\newcommand{\svlamc}{\vlamc}
\newcommand{\svg}{\vg}
\newcommand{\svc}{\vc}
\begin{document}
\pagestyle{empty} \setlength{\footskip}{2.0cm}
\setlength{\oddsidemargin}{0.5cm} \setlength{\evensidemargin}{0.5cm}
\renewcommand{\thepage}{-- \arabic{page} --}
\def\mib#1{\mbox{\boldmath $#1$}}
\def\bra#1{\langle #1 |}      \def\ket#1{|#1\rangle}
\def\vev#1{\langle #1\rangle} \def\dps{\displaystyle}
\def\sla#1{\mbox{$#1\hspace*{-0.17cm}\scriptstyle{/}\:$}}
\renewcommand{\thepage}{-- \arabic{page} --}
   \def\thebibliography#1{\vspace*{0.7cm}\centerline{\large REFERENCES}
     \list{[\arabic{enumi}]}{\settowidth\labelwidth{[#1]}\leftmargin
     \labelwidth\advance\leftmargin\labelsep\usecounter{enumi}}
     \def\newblock{\hskip .11em plus .33em minus -.07em}\sloppy
     \clubpenalty4000\widowpenalty4000\sfcode`\.=1000\relax}\let
     \endthebibliography=\endlist
   \def\sec#1{\addtocounter{section}{1}\section*{\hspace*{-0.72cm}
     \normalsize\bf\arabic{section}.$\;$#1}\vspace*{-0.3cm}}
\vspace*{-1.6cm}\noindent
\hspace*{11.cm}KOBE FHD 01--1\\
\hspace*{11.cm}hep--ph/0103134 \vspace*{0.6cm}\noindent\\
\renewcommand{\thefootnote}{\alph{footnote})}
\centerline{\bf \LARGE Single Diffractive 
\boldmath{$\Lambda_c^+$} Production
in Polarized \boldmath{$pp$}}
\baselineskip=18pt
\centerline{\bf \LARGE Scattering with Polarized Gluon Distribution}
\baselineskip=18pt
\centerline{\bf \LARGE in the Proton}
\vspace*{0.37truein}\\

\begin{center}
{\large
{\sc Kazumasa OHKUMA$^{\:1),\:}$}\footnote{E-mail address:
\tt ohkuma@cnb.phys.ynu.ac.jp},\ \ \ \ 
and \ \ \ \ 
{\sc Toshiyuki MORII$^{\:2),\:}$}\footnote{E-mail address:
\tt morii@kobe-u.ac.jp}\\
\vskip 0.1cm 
\vskip 0.1cm 
%
%
}
\end{center}
{\large
\vspace*{0.4cm}
\vskip 0.2cm
\centerline{\sl $1)$ Faculty of Technology,
Yokohama National University}
\centerline{\sl Hodogaya, Yokohama  240-8501, JAPAN}

\vskip 0.2cm
\centerline{\sl $2)$
 Faculty of Human Development,
Kobe University}
\centerline{\sl Nada, Kobe 657-8501, JAPAN}
}
\vspace*{1.8cm}
\vspace*{0.21truein}

\centerline{ABSTRACT}
\vspace*{0.2cm}
\baselineskip=20pt plus 0.1pt minus 0.1pt
Based on the hard-scattering factorization
which decomposes the diffractive structure function into a pomeron flux 
and a pomeron structure function, 
we study the single diffractive $\Lambda_c^+$ production in polarized
$pp$ scattering:
$p~+~\vp~\rightarrow~p~+~\vlamc~+~X $, 
which will be observed
at forthcoming RHIC experiment.
By analyzing the  cross section
and correlation of the spin polarization between the initial proton 
and produced $\Lambda_c^+$,
we found that the process might be effective for testing
both hard-scattering factorization
and models of the polarized gluon distribution in the proton. \\

\newpage
\renewcommand{\thefootnote}{$\sharp$\arabic{footnote}}
\pagestyle{plain} \setcounter{footnote}{0}
\baselineskip=21.0pt plus 0.2pt minus 0.1pt

\vspace*{12pt}			

As is well known,
the diffractive interaction which 
is characterized by a large rapidity gap event
is described by pomeron exchange  in the Regge theory
\cite{review}.
Although the pomeron carrying the vacuum quantum number
plays a crucial role in diffractive scattering,
nature of the pomeron is still mysterious in the  framework of
quantum chromodynamics (QCD).
Ingelman and Schlein suggested that
the single pomeron exchange in proton-proton interaction
can be probed as a hard scattering
process between a hard parton in the pomeron  emitted  from a proton
and a parton in another proton~\cite{i-s}.
The evidence of the hard partonic structure of the pomeron was first reported
by the UA8 Collaboration at CERN  $p\bar{p}$ collider
at $\sqrt{s}=630$ GeV by observing diffractive dijet production~\cite{AU8}.
This result has been confirmed by further experiments
\cite{h1,zeus,cdf}.
In addition, recent experiment found the
hard-gluon fraction in the pomeron to be $0.7\pm0.2$~\cite{cdf},
which means the pomeron being an almost gluonic object.
Those results suggest us to apply the
factorization theorem to diffractive hard process 
as well as to usual inclusive processes.
However, Collins proved that
the factorization theorem for diffractive deep inelastic scattering (DDIS)
is not expected to be applicable to hadron-hadron collisions,
though this theorem works for the lepton and direct photon
induced hard DDIS~\cite{Coll}.
In fact, the predicted cross sections for hadron-hadron collisions
are several times
larger than the experimental data,
and hence the  hard factorization for DDIS does not work well for
the hadron-hadron collision~\cite{gouli,gouli-monta}.
In order to overcome such difficulty,
Goulianos proposed a phenomenological model in which the structure
of the pomeron is derived from the structure of the parent
hadron~\cite{gouli};
he introduced a renormalized pomeron flux factor which was given
by renormalizing the standard pomeron flux carried by 
a proton to be unity (renormalized pomeron flux model).
Although the  prediction by this model seems to be
in agreement with experimental data~\cite{gouli,gouli-monta}, 
further tests of this model are necessary for  various
processes~\cite{test}.

On the other hand,
the Relativistic Heavy Ion Collider (RHIC) at 
Brookhaven National Laboratory (BNL) 
will start soon.
One of the important purposes of RHIC experiment is to extract  information 
about the  polarized gluon distribution in the proton.
That information is a key to understand the proton spin puzzle
which is still one of the most challenging current topics 
in the nuclear and particle physics~\cite{psp}.
So far there are many models of the polarized parton distribution 
in  the proton, which are  extracted from a fit to the data on 
$g_1(x,Q^2)$~\cite{pdfs,aac}.
Those models can excellently reproduce experimental data on
the polarized structure function of nucleons.
However the behavior of the polarized gluon distributions is quite 
different among these models.
In other words, the data on polarized structure functions of nucleon
and deuteron alone are not enough to distinguish the model of
gluon distribution functions. 
Although various processes have been proposed so far to test the
models of the polarized gluon,  knowledge of it is still poor.

\begin{figure}[htbp] 
\vspace*{13pt}
\centerline{\psfig{file=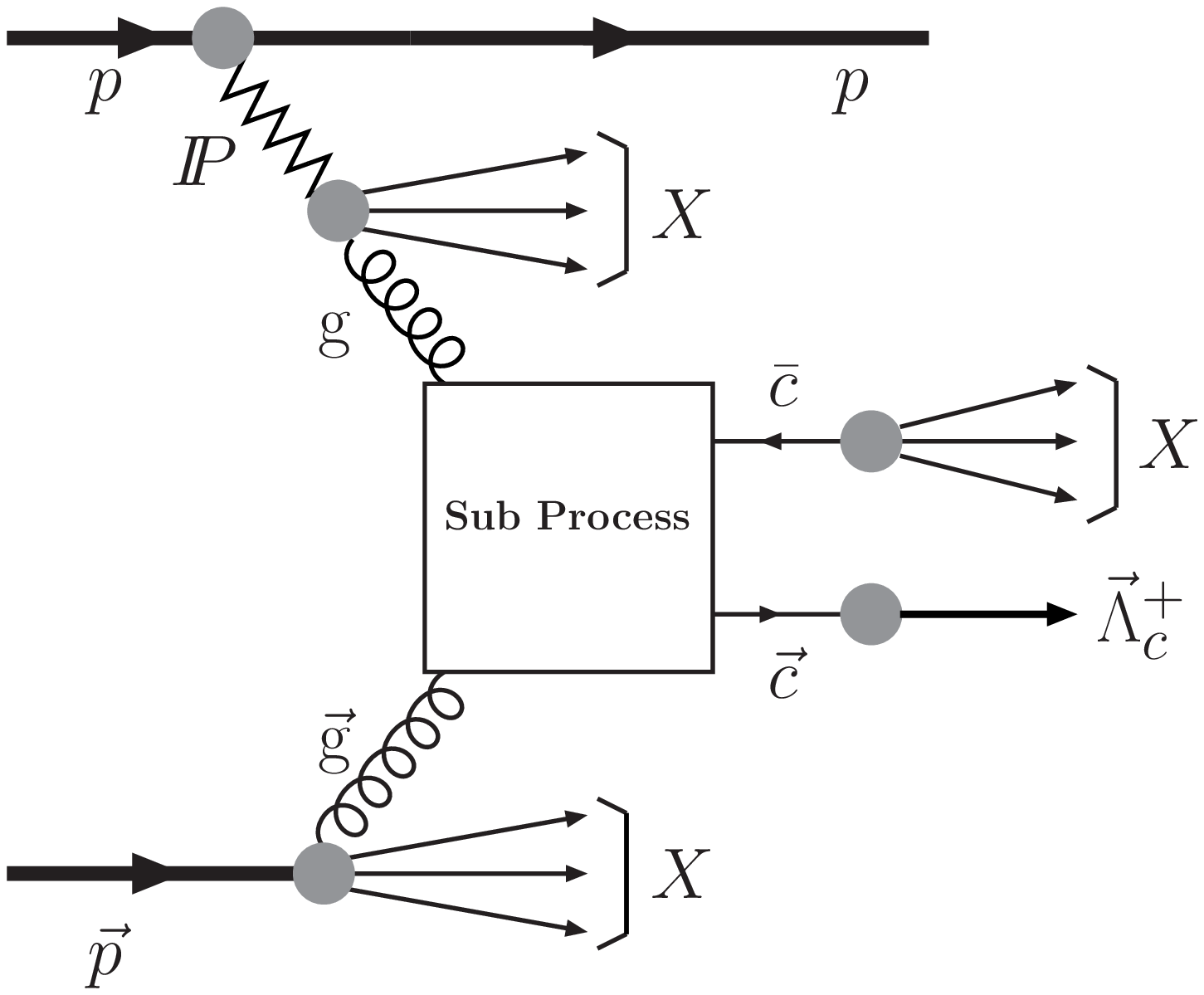,width=2.5TRUEIN,height=2.0TRUEIN }
\psfig{file=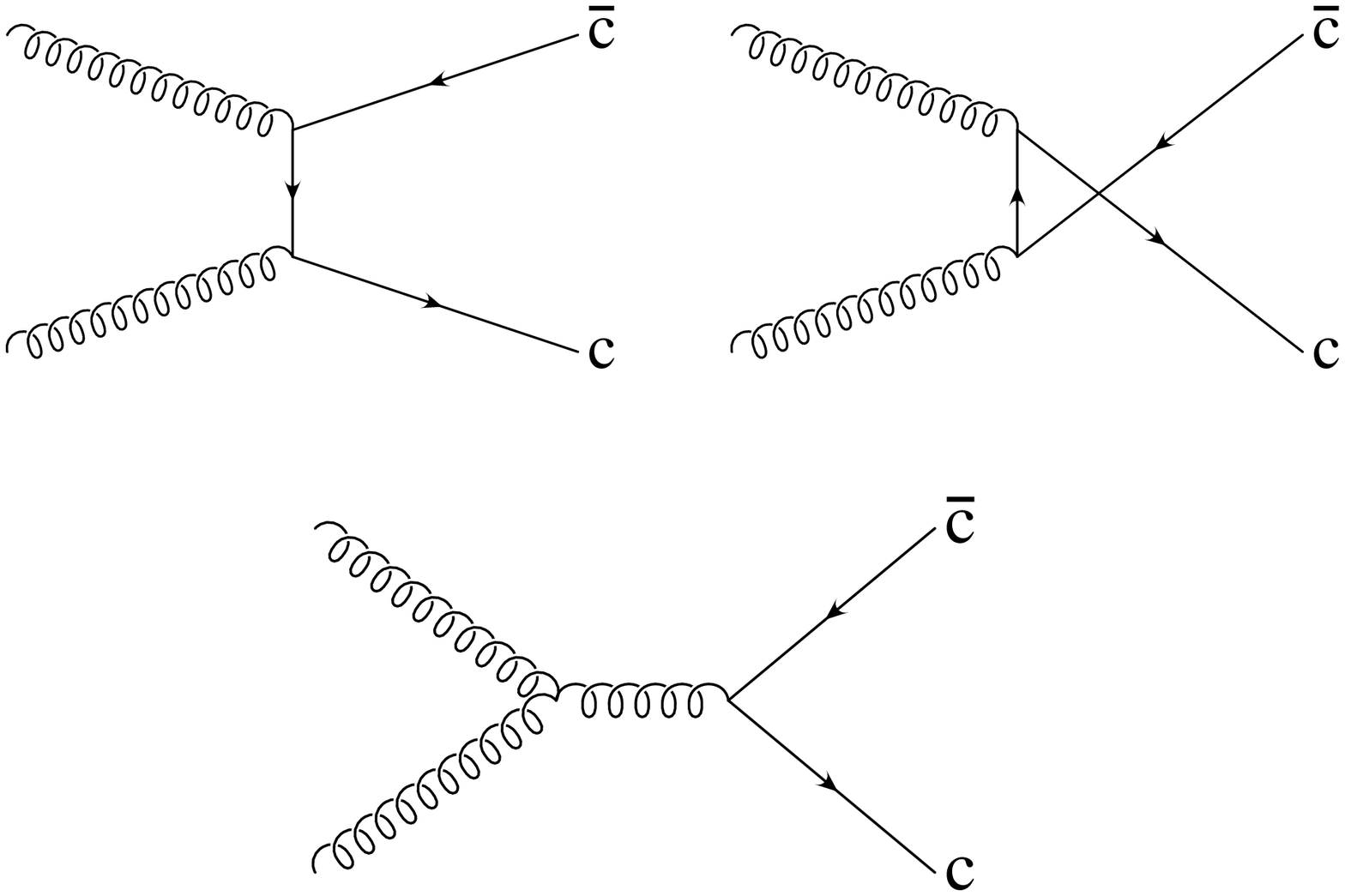,width=2.5TRUEIN,height=2.0TRUEIN }} 
\vspace*{13pt}
\caption{The diagram for 
$p + \vp \rightarrow p + \vlamc + X$
at the  lowest oder in the framework of hard-scattering  factorization
(left side) and its subprocess diagrams (right side).
In the left side, the particles with arrows are polarized.
$I\!\!P$ denotes the pomeron.}
\label{mainpro}
\end{figure}

In this work, we propose another diffractive  semi-inclusive process:
$p + \vp \rightarrow p + \vlamc + X$ ( left side of
Fig.~\ref{mainpro} ),
which can be a test of the renormalized pomeron flux model
and also the polarized gluon distribution function.
In this process, $\Lambda_c^+$ is dominantly produced via
gluon-gluon fusion at the lowest order as
shown in the right side of Fig.~\ref{mainpro}\footnote{
Since the charm quark content is extremely tiny  and
furthermore the pomeron is  dominantly composed of gluons~\cite{cdf},
the gluon-gluon fusion is the dominant process for charm quark pair production.
} ,
where one of the 2 gluons is originated from the pomeron.
Moreover, since $\Lambda_c^+$ is composed of a heavy quark $c$ and
antisymmetrically combined light $u$ and $d$ quarks, the spin of
$\Lambda_c^+$ is expected to be carried by the $c$ quark and thus, 
the production of $\Lambda_c^+$ in polarized proton-proton collisions
gives us
an interesting information about polarized gluons in the initial
proton~\cite{oh-su-mo,ko-mo-ya,oy-mo-ko}.    
To test the hard-scattering factorization and polarized gluon 
distribution model for this attractive process;
\begin{equation}
p~(p_{{}_A}) + \vp~(p_{{}_B}) 
~\rightarrow 
~p~(p'_{{}_A}) + \vlamc~(p_{{}_{\Lambda_{{}_c}^+}}) +~X,
\label{m-pro}
\end{equation}
whose lowest order subprocess
is\footnote{
A subprocess in this model is a hard gluon-gluon scattering
in the pomeron-proton system, as shown  in Fig.~\ref{mainpro}.
}
\begin{equation}
{\rm g}(p_{{}_a})+\vg (p_{{}_b}) \rightarrow 
\vc(p_c) + \bar{c}(p_{\bar{c}}) 
\label{sub-pro},
\end{equation}
where $p_i~(i=A, B, a, b, c, \bar{c},\Lambda_c^+)$ in the parentheses
denote the four-momentum of each particle
and the over-arrow means that an initial gluon with momentum $p_b$ and 
a produced $c$ quark with momentum $p_c$ are polarized,
we introduce two useful observables;
one is  the $d \Delta \sigma/d p_{{}_T}$ defined in Eq.(\ref{dsigma}),
which we call hereafter the polarized differential cross section,
and
the other is  the $A_{LL}$ defined in Eq.(\ref{ALL}), which we call
the spin correlation asymmetry;
\begin{eqnarray}
&&\frac{d \Delta\sigma}{dp_{{}_T}}
\equiv \frac{
d\sigma(++) - d\sigma(+-) + d\sigma(--) - d\sigma(-+)}
{dp_{{}_T}},\label{dsigma} \\
&&A_{LL} \equiv 
\frac{ \left[d\sigma(++) - d\sigma(+-) + d\sigma(--) - d\sigma(-+)\right]
/d p_{{}_T}}
     { \left[d\sigma(++) + d\sigma(+-) + d\sigma(--) + d\sigma(-+)\right]
/d p_{{}_T}}\nonumber \\
&&\phantom{A_{LL}}\equiv \frac{d \Delta \sigma/dp_{{}_T}}{d \sigma/dp_{{}_T}},
\label{ALL}
\end{eqnarray}
where $d\sigma(+-)/d p_{{}_T}$, 
for example, denotes the spin-dependent differential
cross section with  positive helicity of the target proton and
negative helicity of the produced $\Lambda_c^+$.

 Let us consider the process in the proton-proton center-of-mass  frame.
In this frame, we can take four-momenta of $p_i$ as follows;
\begin{eqnarray}
&&p_{{}_{A,B}}=\frac{\sqrt{s}}{2}(1,\mp \beta ,\vzero)\ \ {\rm with} \ \  
\beta \equiv \sqrt{1-\frac{4m^2_{p}}{s}}, \nonumber \\
&&p_{{\Lambda_c^+}}=(E_{\Lambda_c^+},p_{{}_L},\vp_{{}_T})\nonumber \\
&& \phantom{p_{\lambda}}
=(\sqrt{m_{\Lambda_c^+}^2 + p_{{}_T}^2 {\rm cosec} ^2 \Theta},
p_{{}_T}\cot \Theta, \vec p_{{}_T}),\nonumber\\
&&p_{{}_{I\!\!P}}=\xi p_{{}_A},
\ \ \ 
p_{{}_{a}}=x_{{}_{a}}p_{{}_{I\!\!P}},\ \ \ 
p_{{}_{b}}=x_{{}_{b}}p_{{}_{B}},
\ \ \ \ p_{{}_c}=\frac{p_{{\Lambda_c^+}}}{z},
\label{kine}
\end{eqnarray}
where the first, second and third components 
in parentheses are the energy,
the longitudinal momentum and the transverse momentum, respectively.
$m_i$ is  the mass of the $i$-particle.
The polarized differential cross section,
${d \Delta \sigma}/{d p_{{}_T}}$, can be calculated as follows;
\begin{eqnarray}
\frac{d \Delta \sigma}{d p_{{}_T}}&=&
\int^0_{-1} dt
\int^{\xi^{\rm max}}_{\xi^{\rm min}} d\xi
\int^{\Theta^{\rm max}}_{\Theta^{\rm min}} d\Theta
\int^{1}_{x_a^{\rm min}} dx_{{}_a}
\int^{1}_{x_b^{\rm min}} dx_{{}_b} \nonumber \\
&&\times f^{\rm RN}_{I\!\!P/p}(\xi,t)f_{{\rm g}/I\!\!P}(x_a,Q^2)
\Delta G_{\svg/\svp}(x_b,Q^2)\nonumber\\
&&\times \frac{d \Delta \hat{\sigma}}{d \hat{t}}
{\cal J}
\Delta {\rm D}_{\svlamc/\svc}(z), 
\label{tcross}
\end{eqnarray}
where
$t$ is the square of the four-momentum transfer of the proton
which  emits the pomeron.
${\cal J}$ is the Jacobian which transforms the variables 
$z$  and $\hat{t}$ into  $\Theta$ and $p_{{}_T}$. 
$f_{{\rm g}/I\!\!P}(x_{{}_a},Q^2)$ 
and
$f^{\rm RN}_{I\!\!P/p}(\xi,t)$ 
are the hard gluon distribution function in the pomeron and
the renormalized pomeron flux in the proton, respectively.
The renormalized pomeron flux 
which was proposed by Goulianos 
to predict the observed single
diffractive cross section~\cite{gouli},
is defined by
\begin{equation}
f^{\rm RN}_{I\!\!P/p}(\xi,t) \equiv D f_{I\!\!P/p}(\xi,t), \nonumber \\
\end{equation}
where $f_{I\!\!P/p}(\xi,t)$ and $D$ are
the standard pomeron flux and renormalization factor, respectively.
$f_{I\!\!P/p}(\xi,t)$ is  given by
\begin{equation}
f_{I\!\!P/p}(\xi,t)=K \xi^{1-2\alpha(t)}F^2(t),
\label{flux}
\end{equation}
with the parameters which are chosen as~\cite{gouli}
\begin{eqnarray}
K=0.73~{\rm GeV}^2 ,&~~~&\alpha(t)=1+0.115+0.26 [ {\rm GeV^{-2}}] t,
 \nonumber\\
&&F^2(t)=e^{4.6t}.\nonumber
\end{eqnarray}

The renormalization factor $D$ is defined as \cite{gouli}
\begin{equation}
      D={\rm min}(1,\frac{1}{N})
\end{equation}
with
\begin{eqnarray}
N & =& \int^{0.1}_{M^2_0/s} d\xi
             \int_{t=0}^{t=\infty} f_{I\!\!P/p}(\xi,t)dt
\label{n-factor}
\end{eqnarray}
where   $M_0^2$ is 1.5 ${\rm GeV}^2$ being the 
effective diffractive threshold 
and the upper limit for the $\xi$ integration is a coherence limit.
In addition,
$\Delta G_{\svg/\svp}(x_b,Q^2)$
and $\Delta {\rm D}_{\svlamc/\svc}(z)$  
represent the polarized gluon
distribution function in the proton  and 
the polarized  fragmentation function of the outgoing charm quark 
decaying into a polarized $\vlamc$, respectively.

Since the subprocesses considered here are the same with the ones
discussed in Ref.[14], here we repeat some important
formulas for reader's convenience.
By using the kinematical variables in Eq.(\ref{kine}),
the polarized differential cross section,
$d \Delta \hat{\sigma}/d \hat{t}$,
for the subprocess
is calculated to be
\begin{eqnarray}
\frac{d \Delta\hat{\sigma}}{d \hat{t}}
&=&\frac{\pi \alpha^2_s}{\hat{s}}
\left[ \frac{m_c^2}{24} \left\{
\frac{9 \hat{t}_1 -19 \hat{u}_1}{\hat{t}_1 \hat{u}_1}
+\frac{8 \hat{s}}{\hat{u}_1^2}\right\} \right.\nonumber \\
&&\left. \phantom{-\frac{\pi \alpha^2_s}{\hat{s}}}
+\frac{\hat{s}}{6}
\left\{
\frac{\hat{t}_1-\hat{u}_1}{\hat{t}_1 \hat{u}_1}
\right\}
-\frac{3}{8}\left\{ \frac{2 \hat{t}}{\hat{s}} +1 \right\}
\right],
\label{cross}
\end{eqnarray}
where
$\hat{s}$,~ $\hat{t}_1$ and $\hat{u}_1$ are defined as
$\hat{s} \equiv (p_{{}_a}+p_{{}_b})^2$,
$\hat{t}_1 \equiv (p_{{}_b}-p_{{}_c})^2 - m_c^2 $
and
$\hat{u}_1 \equiv (p_{{}_a}-p_{{}_c})^2 - m_c^2 $,
respectively.
The Jacobian ${\cal J}$ is given by
\begin{equation}
{\cal J}=\frac{2s\beta p_{{}_T}^2 {\rm cosec}^2 \Theta}
{z (s - 2 m_p^2) \sqrt{m^2_{\Lambda_c^+}+p_{{}_T}^2 {\rm cosec}^2 \Theta}},
\label{jabocian}
\end{equation}
where  $s$ and $z$ are defined as
$
s \equiv (p_{{}_A}+ p_{{}_B})^2 
$
and
$
z\equiv \frac{x_1}{x_a \xi} + \frac{x_2}{x_b}
$
, respectively,
with $x_1 \equiv \frac{2 p_{{}_B} p_{{}_{\Lambda_c}}}{s-2m^2_p}$,
$x_2 \equiv \frac{2 p_{{}_A} p_{{}_{\Lambda_c}}}{s-2m^2_p}$.

In order to estimate the asymmetry, $A_{LL}$, we need the unpolarized cross
section which can be obtained by replacing polarized functions,
$\Delta G_{\svg/\svp},~d \Delta \hat{\sigma}/d \hat{t}$
and $\Delta {\rm D}_{\svlamc/\svc}$ by unpolarized functions,
$G_{{\rm g}/p},~d \hat{\sigma}/d \hat{t}$ and 
${\rm D}_{\Lambda_c^+/c}$, respectively, in Eq. (\ref{tcross}).
The explicit formula of 
the unpolarized differential cross section for this subprocess
was given by Babcock et al. \cite{unpol}.
As for the gluon distributions, we take the AAC~\cite{aac} 
and the GRSV01~\cite{grsv} parameterization models for the polarized
distribution and the GRV98~\cite{grv} model for 
the unpolarized one. 
Here, we set $Q^2$ as $(2m_c)^2$ for each models.
As for the unpolarized fragmentation function, 
we use Peterson fragmentation function~\cite{peter}, 
$D_{c \to \Lambda_c^+}(z)$.
However, unfortunately at present there are no established 
polarized fragmentation function $\Delta D_{c\to \Lambda_c^+}$ because 
of lack of experimental data.  Thus, by analogy with the study on $\Lambda$
polarization~\cite{deF}, we take the following ansatz:
$$
\Delta D_{\vec{c}\to \vec{\Lambda}_c^+}(z)= C_{c\to \Lambda_c^+} 
D_{c \to \Lambda_c^+},
$$
where $C_{c \to \Lambda_c^+}$ is scale-independent spin transfer coefficient.
Here we apply the analysis on $\Lambda$ production to $\Lambda_c^+$
production and choose the following two typical models:
\begin{quote} 
(i) $C_{c\to \Lambda_c^+}=1$ (non-relativistic quark model), \\ 
(ii) $C_{c\to \Lambda_c^+}=z$ (Jet fragmentation model~\cite{Bartl}).
\end{quote}
For the hard gluon distribution function of the pomeron, we use
\cite{cdf,buni-ing};
\begin{equation}
x f_{{\rm g}/I\!\!P}(x,Q^2)=f_{\rm g} 6x(1-x),\ \ \ f_{\rm g}=0.7\pm 0.2,
\end{equation}
where $f_{\rm g}$ is the hard gluon fraction in the pomeron.
To determine the value of the renormalization factor $D$ given by Eq.(7),
we estimated the value of $N$ defined by Eq.(\ref{n-factor}) to be
$
N = 3.4~(5.0)
$
for  $\sqrt{s}=$200 (500) GeV
using Eq.(\ref{flux})
\cite{gouli-monta}.

For numerical calculation, we use, as input parameters,
$m_c = 1.25$ GeV, $m_p = 0.938$ GeV and 
$m_{{}_{\Lambda_c^+}} = 2.285$ GeV~\cite{pd}.
In numerically integrating Eq. (\ref{tcross}), the minimum 
values of $\xi,x_{{}_a}$ and $x_{{}_b}$ are given by
$
\xi^{\rm min} =\frac{x_1}{1-x_2},
\ \ x^{\rm min}_{{}_a}=\frac{x_1}{\xi(1-x_2)}~{\rm and}
\ \ x^{\rm min}_{{}_b}=\frac{\xi x_{{}_a}x_{{}_2}}{\xi x_{{}_a}-x_1},
$
respectively.
In addition, to reduce possible non-pomeron contribution,
we set $\xi^{\rm max}$=0.05 as usual~\cite{gouli,gouli-monta,test}.
Furthermore, in order to get rid of
the produced $\Lambda_c^+$ entering the beam pipe,
we limit the integration region of $\Theta$ for produced $\Lambda_c^+$  as
$\frac{\pi}{6} \leq \Theta \leq \frac{5\pi}{6}$.  
 $p_{{}_T}$ region is kinematically  constrained from the condition of the
diffractive production.
\begin{figure}[htbp]
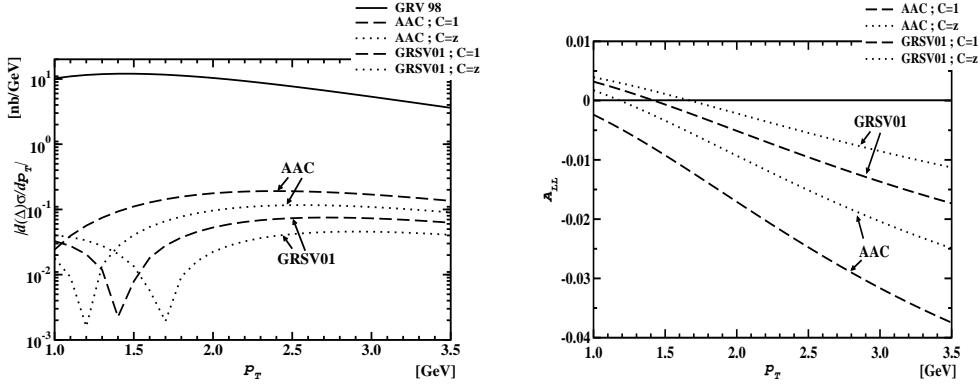
 
\vspace*{13pt}
\centerline{\psfig{file=200cross.eps,width=2.5TRUEIN,height=2.0TRUEIN }
\hspace*{0.5cm}
\psfig{file=200all.eps,width=2.3TRUEIN,height=2.0TRUEIN }} 
\vspace*{13pt}
\caption{
The unpolarized and polarized differential cross
 section (left panel) and the spin correlation asymmetry (right panel) 
as a function of 
$p_{{}_T}$ at $\sqrt{s}=200$ GeV.
The solid line in the left panel represents the unpolarized
differential cross section with the GRV98 model for the unpolarized
gluon distribution.
The long-dashed and dashed lines show  
the polarized differential cross section calculated with
$C_{c\to \Lambda_c^+}=1$ and $C_{c\to \Lambda_c^+}=z$, respectively, 
for AAC and GRSV01 models of polarized gluon distributions.
The value of ${d \Delta \sigma}/{d p_{{}_T}}$ is negative for $p_{{}_T}$
region larger than the value
corresponding to the apparent sharp dip.
The same combination of the models for polarized gluon distributions and 
polarized fragmentation functions is adopted in the right panel.
}
\label{200fig}
\end{figure}
\begin{figure}[htbp]
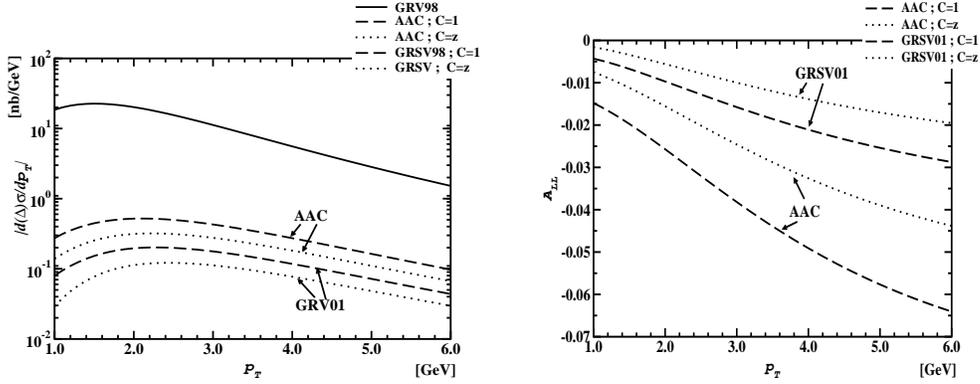
 
\vspace*{13pt}
\centerline{\psfig{file=500cross.eps,width=2.5TRUEIN,height=2.0TRUEIN }
\hspace*{0.5cm}
\psfig{file=500all.eps,width=2.3TRUEIN,height=2.0TRUEIN }} 
\vspace*{13pt}
\caption{
The same as in Fig. \ref{200fig}, but for $\sqrt{s}=500$ GeV.
}
\label{500fig}
\end{figure}

We show the $p_{{}_T}$ distribution of
$|d (\Delta) \sigma /d p_{{}_T}|$ and 
$A_{LL}$ in Fig.~\ref{200fig} for $\sqrt{s} = 200$ GeV  
and in Fig.~\ref{500fig} for
$\sqrt{s} = 500$ GeV, respectively.
Notice that the absolute value of  $d \Delta \sigma /d p_{{}_T}$
is presented in these figures,
 because the negative value of 
$d \Delta \sigma /d p_{{}_T}$ cannot be depicted in the figure
which has an ordinate with logarithmic scale.
Actually, the value of $d \Delta \sigma /d p_{{}_T}$ is
negative for $p_{{}_T}$ region larger than the value corresponding to 
the apparent sharp dip shown in Fig.~\ref{200fig}.
Information on the dip is expected to be useful for 
distinguishing the models of 
the polarized gluon distribution functions because each parameterization
model has a dip at different $p_{{}_T}$ as shown in the left panel
of Figs.~\ref{200fig}.  At $\sqrt{s}=500~$GeV, the dip is not seen as
shown in the left panel of Fig.~\ref{500fig}, because the dip is in the
region smaller than $p_{{}_T}=1~$GeV for the case of $\sqrt{s}=500~$GeV.
Thus, the polarized differential cross section are actually negative
in the kinematical region presented in Fig.~\ref{500fig}.
From the right panel of  Figs.~\ref{200fig} and \ref{500fig},
it seems that
the $A_{LL}$ is an effective observable
to distinguish various models of polarized gluon distribution functions,
even if there are uncertainties of the spin-dependent fragmentation 
function.
Notice that 
if the renormalized pomeron flux model is not taken into account for
the present process,
the polarized differential cross section and
unpolarized differential cross section become
3.4 (5.0) times larger than our calculation 
for $\sqrt{s}=200(500)$ GeV. 
Therefore, measurement of those cross sections
can be a good test of the renormalized pomeron flux model,
though  the  $A_{LL}$ is not useful for testing the renormalized flux model
because the renormalized flux factors are canceled out between the numerator
and the denominator of Eq.(\ref{ALL}).

Finally, some comments are in order, regarding uncertainty of the calculated
result on the value of $m_c$ and the choice of $Q^2$. 
We have examined the variation of $A_{LL}$ on $m_c$ for the region of
1.15GeV $\leq~ m_c~\leq$ 1.35GeV and found that the result remain 
to be unchanged. 
We also found that the $Q^2$ dependence is rather insensitive;
we have examined the 3 cases, 
(i)  $Q^2=(2 m_c)^2$,
(ii) $Q^2=p_{{}_T}^2$ 
and
(iii) $Q^2=m_{\Lambda_c^+}^2+p_{{}_T}^2$,
 without much difference.

In summary,
we have calculated the polarized differential cross 
section, ${d \Delta \sigma}/{d p_{{}_T}}$,
and the spin correlation asymmetry, $A_{LL}$,
for the single diffractive $\Lambda_c^+$ production in polarized 
$pp$ reactions 
at $\sqrt{s}=200$ GeV and
$\sqrt{s}=500$ GeV, based on the hard-scattering
factorization with the renormalized pomeron flux.
We found that
${d \Delta \sigma}/{d p_{{}_T}}$
and  $A_{LL}$ largely depend on the model
of the polarized gluon distribution function.
Therefore, the process looks promising for testing the models
of polarized gluon distribution function.
Moreover, it is expected that
the measurement of ${d (\Delta) \sigma}/{d p_{{}_T}}$ is quite effective 
for testing the renormalized  pomeron flux model.
In order to get more reliable  prediction,
the next-to-leading order calculation and error estimation
might be  necessary, which will be given in the forthcoming paper.
In addition, since our prediction somewhat depends  on the model of
polarized fragmentation function, further experimental and theoretical
study on polarized fragmentation functions is necessary for 
getting more reliable predictions.

Although the present calculation is confined in the leading order,
the results are interesting and 
we hope our prediction will be tested in the forthcoming RHIC
experiment.

\vspace*{0.4cm}
The authors thanks N. Saito for informing the physical parameters
and condition of the RHIC experiment.
We are also deeply grateful  to M.~Reya
and W.~Vogelsang for sending Fortran program of GRSV01.
One of the authors (T.M.) would like to thank for the financial
support by the Grant-in-Aid for Scientific Research,
Ministry of Education, Science and Culture, Japan (No.11694081).

\end{document}